\documentclass[superscriptaddress,preprintnumbers,amsmath,amssymb,twocolumn,floats]{revtex4-1}
\usepackage{txfonts}
\usepackage{amssymb}
\usepackage{graphicx}
\usepackage{color}
\definecolor{RED}{rgb}{1,0,0}
\definecolor{BLUE}{rgb}{0,0,1}
\usepackage{sidecap}
\usepackage{CJK}
\usepackage{txfonts}
\usepackage{url}
\begin{document}

\title{Electronic structure and 4$f$-electron character in Ce$_2$PdIn$_8$ \\ studied
by angle-resolved photoemission spectroscopy}


\author{Q. Yao}
\affiliation{State Key Laboratory of Surface Physics, Department of Physics, and Laboratory of Advanced Materials, Fudan University, Shanghai 200438, China}
\affiliation{State Key Laboratory of Functional Materials for Informatics, Shanghai Institute of Microsystem and Information Technology (SIMIT), Chinese Academy of Sciences, Shanghai 200050, China}

\author{D. Kaczorowski}
\affiliation{Institute of Low Temperature and Structure Research, Polish Academy of Sciences, P. O. Box 1410, 50-950 Wroclaw, Poland}
\affiliation{Centre for Advanced Materials and Smart Structures, Polish Academy of Sciences, Okolna 2, 50-422 Wroclaw, Poland}

\author{P. Swatek}
\affiliation{Institute of Low Temperature and Structure Research, Polish Academy of Sciences, P. O. Box 1410, 50-950 Wroclaw, Poland}
\affiliation{Division of Materials Science and Engineering, Ames Laboratory, Ames, Iowa 50011, USA}
\affiliation{Department of Physics and Astronomy, Iowa State University, Ames, Iowa 50011, USA}
\author{D. Gnida}
\affiliation{Institute of Low Temperature and Structure Research, Polish Academy of Sciences, P. O. Box 1410, 50-950 Wroclaw, Poland}

\author{C. H. P. Wen}
\author{X. H. Niu}
\author{R. Peng}
\author{H. C. Xu}
\affiliation{State Key Laboratory of Surface Physics, Department of Physics, and Laboratory of Advanced Materials, Fudan University, Shanghai 200438, China}

\author{P. Dudin}

\affiliation{Diamond Light Source, Harwell Science and Innovation Campus, Didcot OX11 0DE, United Kingdom}


\author{S. Kirchner}
\affiliation{Zhejiang Institute of Modern Physics, Zhejiang University, Hangzhou, Zhejiang 310027, China}

\author{Q. Y. Chen}
\email{sheqiuyun@126.com}
\affiliation{Science and Technology on Surface Physics and Chemistry Laboratory, Mianyang 621908, China}

\author{D. W. Shen}
\email{dwshen@mail.sim.ac.cn}
\affiliation{State Key Laboratory of Functional Materials for Informatics, Shanghai Institute of Microsystem and Information Technology (SIMIT), Chinese Academy of Sciences, Shanghai 200050, China}
\affiliation{CAS Center for Excellence in Superconducting Electronics (CENSE), Shanghai 200050, China}

\author{D. L. Feng}

\affiliation{State Key Laboratory of Surface Physics, Department of Physics, and Laboratory of Advanced Materials, Fudan University, Shanghai 200438, China}
\affiliation{Collaborative Innovation Center of Advanced Microstructures, Nanjing 210093, China}

\begin{abstract}
The localized-to-itinerant transition of $f$ electrons lies at the heart of heavy fermion physics, but has only been directly observed in single-layer Ce-based materials. Here we report a comprehensive study on the electronic structure and nature of the Ce 4$f$ electrons in the heavy fermion superconductor Ce$_2$PdIn$_8$, a typical $n$=2 Ce$_n$M$_m$In$_{3n+2m}$ compound, using high-resolution and 4$d$-4$f$ resonant photoemission spectroscopies. The electronic structure of this material has been studied over a wide temperature range, and hybridization between $f$ and conduction electrons can be clearly observed to form Kondo resonance near the Fermi level at low temperature. The characteristic temperature of the localized-to-itinerant transition is around 120~K, which is much higher than its coherence temperature $T_{coh} \sim30$\,K.
\end{abstract}


\maketitle


In heavy fermion materials, $f$ electrons generally exhibit both itinerant and localized feature as the temperature changes. According to the standard Kondo lattice model, $f$ electrons are localized at high temperatures, while conduction electrons would screen the local moments of $f$ electrons to form Kondo singlets upon cooling, eventually resulting in an itinerant heavy Fermi-liquid ground state~\cite{coleman}. On the other hand, the recent two-fluid theory suggests that part of an atom's $f$ electrons give rise to the itinerant properties at low temperatures, while others account for the localization behavior at higher temperatures~\cite{yang1,yang2,yang3,yang4}. Experimentally, detailed evolution of the localized-to-itinerant transition has also been hotly debated~\cite{f1, f2, f3, f4,CeCoIn5_qiuyun}. Recently, the hybridization between $f$ and conduction electrons were extensively studied, and direct evidence for such localized-to-itinerant transition has been observed in the Ce$M$In$_5$ family ($M$=Co, Rh, Ir)~\cite{CeRhIn5_ARPES, CeCoIn5_qiuyun, CeIrIn5_ARPES2,CeRhIn5_qiuyun,CeIrIn5_qiuyun}. However, in other Ce-based heavy-fermion compounds, temperature evolution of the localized-to-itinerant transition of the $f$ electrons has never been reported to the best of our knowledge~\citep{CeCoIn218_ARPES,CeRhIn218_ARPES}.

\begin{figure*}
\includegraphics[width=\textwidth]{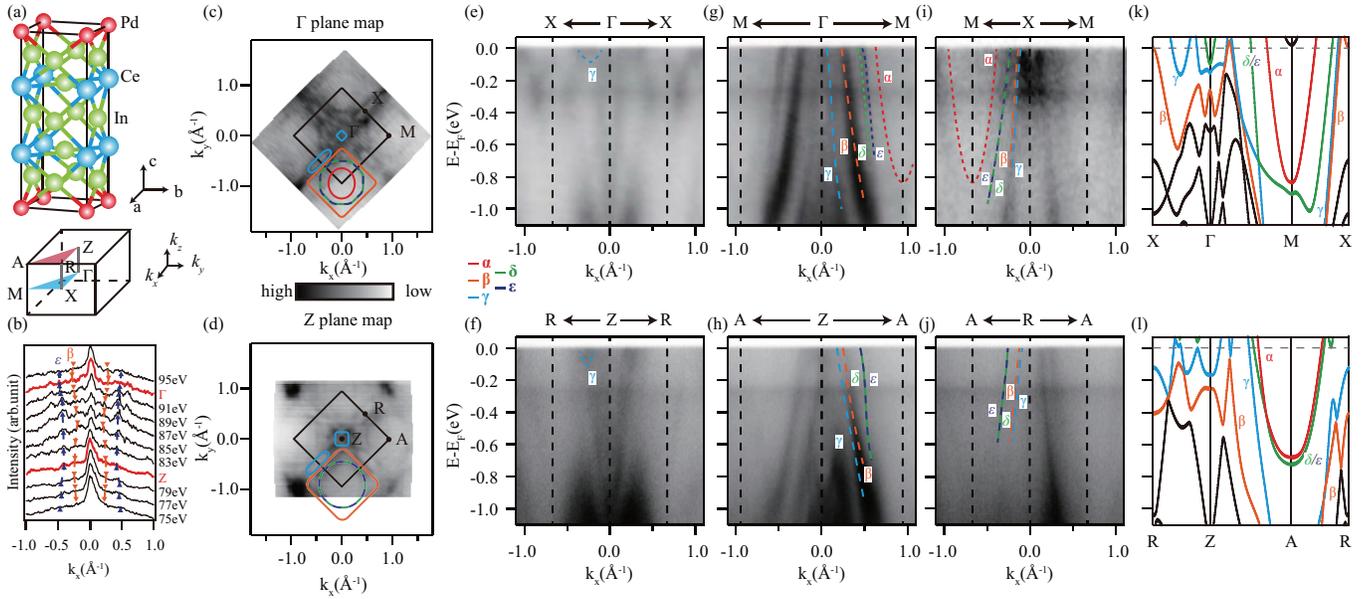}
\caption{(a) Crystal structure and the three-dimensional Brillouin zone of tetragonal Ce$_2$PdIn$_8$. (b) Momentum distribution curves taken with 75$\sim$95 eV photons along $M$-$\Gamma$-$M$. (c)-(d) Photoemission intensity $\Gamma$ and $Z$ maps of Ce$_2$PdIn$_8$ at $E_F$, taken with 80 eV and 93 eV photon energies, respectively. The intensity was integrated over ($E_F$-10meV, $E_F$+10meV). (e), (g) and (i) Photoemission intensity plots along the \emph{M}-\emph{$\Gamma$}-\emph{M}, \emph{X}-\emph{$\Gamma$}-\emph{X} and \emph{M}-\emph{X}-\emph{M} directions. (f), (h) and (j) Photoemission intensity plots along the \emph{A}-\emph{Z}-\emph{A}, \emph{R}-\emph{Z}-\emph{R} and \emph{A}-\emph{R}-\emph{A} directions, respectively. (k-l) Calculated bands along high-symmetry directions, which have been shifted upwards by $\sim$150 meV to match the experimental results. }
\label{bands}
\end{figure*}

The Ce$_n$$M$In$_{3n+2}$ ($M$=Co, Rh or Ir, with $n$=1, 2, $\infty$) heavy-fermion compounds consist of $n$ layers of CeIn$_3$, stacked sequentially along the $c$ axis with one intervening layer of $M$In$_2$ \cite{Bao.01}. The electronic structure of the  $n$=$\infty$ and 1 compounds, CeIn$_3$ and Ce$M$In$_5$, has been extensively studied by angle-resolved photoemission spectroscopy (ARPES)~\cite{zhangyun_CeIn3,CeIn3_ARPES,CeRhIn5_ARPES, CeCoIn5_qiuyun, CeIrIn5_ARPES2}, which has been proved to be a powerful tool to directly observe the behavior of $f$ electrons. For the $n$=2 compounds, Ce$_2$CoIn$_8$ and Ce$_2$RhIn$_8$ have been investigated through off-resonant ARPES, and Ce 4$f$ electrons were found to be predominantly localized in both compounds~\cite{CeCoIn218_ARPES,CeRhIn218_ARPES}. However, to date there have been no reports on these $n$=2 compounds by means of on-resonance ARPES, in which the use of the 4$d$-4$f$ resonance would largely enhance the $f$-electron photoemission matrix element. Moreover, for another important $n$=2 member, Ce$_2$PdIn$_8$ with the unique paramagnetic ground state, de Haas-van Alphen (dHvA) measurements \cite{Ce218_dHvA} and X-ray photoelectron spectroscopy (XPS) \cite{Ce218_XPS} both suggest  delocalized character of the Ce 4$f$ states therein, yet a direct observation of the $f$-electron behavior, and particularly the itinerant-to-localized transition, in this compound is still lacking.

In this Rapid Communication, we present a systematic ARPES study on Ce$_2$PdIn$_8$. Its electronic structure is comprehensively revealed and directly compared with first-principles calculations. In particular, the character of the $f$ electrons is studied by resonant ARPES, and the localized-to-itinerant transition is directly found to set in at a surprisingly high temperature.

High-quality single crystals of Ce$_2$PdIn$_8$ were synthesized by a self-flux method as described in Ref.~\cite{Ce218Kondo1}. ARPES measurements were conducted with photons from Beamline I05 of the Diamond light source, which is equipped with a Scienta R4000 electron analyzer. The overall energy resolution is 15$\sim$25~meV depending on the photon energy, and the angular resolution was set to 0.3\,$^\circ$. Samples were cleaved at 10~K and then measured between 10 and 190~K under a vacuum better than 5 $\times$10$^{-11}$ Torr. Electronic band structure calculations of Ce$_2$PdIn${_8}$ were performed with the all-electron general potential linearized augmented plane-wave (FL-APW) method as implemented in the Wien2k code~\cite{wien2k}. The exchange and correlation effects were treated using generalized gradient approximation (GGA) in the form proposed by Perdew, Wang and Ernzerhof~\cite{GGA}. The Ce 2$f$ electrons were removed from valence band as it was previously done in DFT calculations for Ce$_2$RhIn$_8$
~\cite{CeRhIn218_ARPES, CeIrIn5_ARPES2}. Spin-orbit coupling (SOC) was included as a second variational step with a basis of scalar-relativistic eigenfunctions, after the initial calculation was converged to self-consistency. The Monkhorst--Pack~\cite{MK} special $k$-point scheme with $23\times23\times9$ mesh was used resulting in 390 $k$ points in the irreducible part of the Brillouin zone. The wave functions in the interstitial region were expanded using plane waves with a cutoff of  ($R_{mt}K_{max}$) = 8, where $K_{max}$ is the plane wave cut-off, and $R_{MT}$ is the smallest of all atomic sphere radii. Experimental lattice parameters have been used with atoms fixed in their bulk positions.

\begin{figure}[t]
\includegraphics[width=\columnwidth]{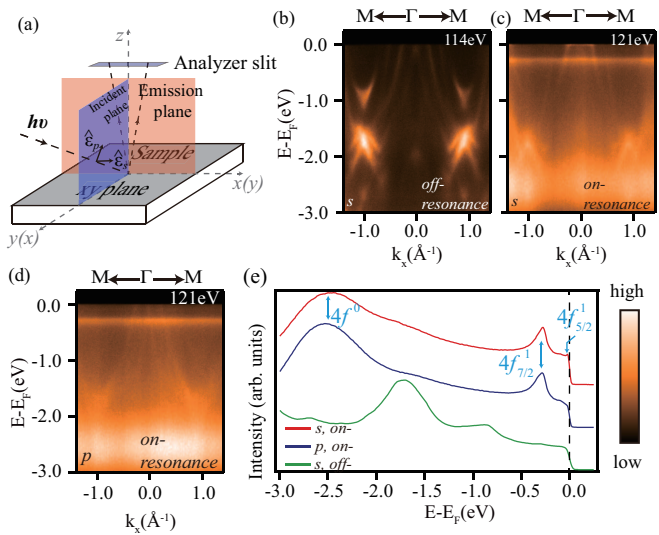}
\caption{(a) Experimental geometry for polarization-dependent ARPES. The emission plane is defined by the analyzer slit and the sample surface normal. $p$- and $s$-polarized light have polarization in and out of the emission plane, respectively. (b)-(c) Photoemission intensity plots along $M$-$\Gamma$-$M$, taken with off-resonance (114 eV) and on-resonance (121\,eV) $s$-polarized light, respectively. (d) On-resonance (121\,eV) photoemission intensity plot along \emph{M}-\emph{$\Gamma$}-\emph{M}, using $p$-polarized light. (e) Angle-integrated EDCs of panels (b)-(d). Data were taken at 11 K.}
\label{resonance}
\end{figure}


Fig.~\ref{bands}(a) illustrates the crystal structure of Ce$_2$PdIn$_8$ and the corresponding three-dimensional Brillouin zone (BZ). This compound forms in the tetragonal Ho$_2$CoIn$_8$-type crystal structure ($P4/mmm$) with lattice parameters $a$=$b$=4.69\,\AA\ and $c$=12.185\,\AA, and consists of a sequence of two CeIn$_3$ layers alternating with a PdIn$_2$ layer along the $c$ axis~\cite{CePdIn_crystal}. To investigate its three-dimensional electronic structure, we have performed detailed $k_z$-dependent measurements by varying the photon energy between 75 and 95~eV. As shown in Fig.~\ref{bands}(b), there exists weak photon-energy dependence of peak positions in the momentum distribution curves (MDCs) taken along the high symmetry $\Gamma$-$M$ direction, indicating the quasi-two-dimensional character. An inner potential of 13~eV, comparable to that of CeIn$_3$~\cite{CeIn3_ARPES}, has been obtained through the best fit to the periodic variation of the MDCs' peak positions according to the free-electron final-state model~\cite{kz}, allowing us to determine the photon energies corresponding to the high-symmetry planes.

\begin{figure*}
    \includegraphics[width=\textwidth]{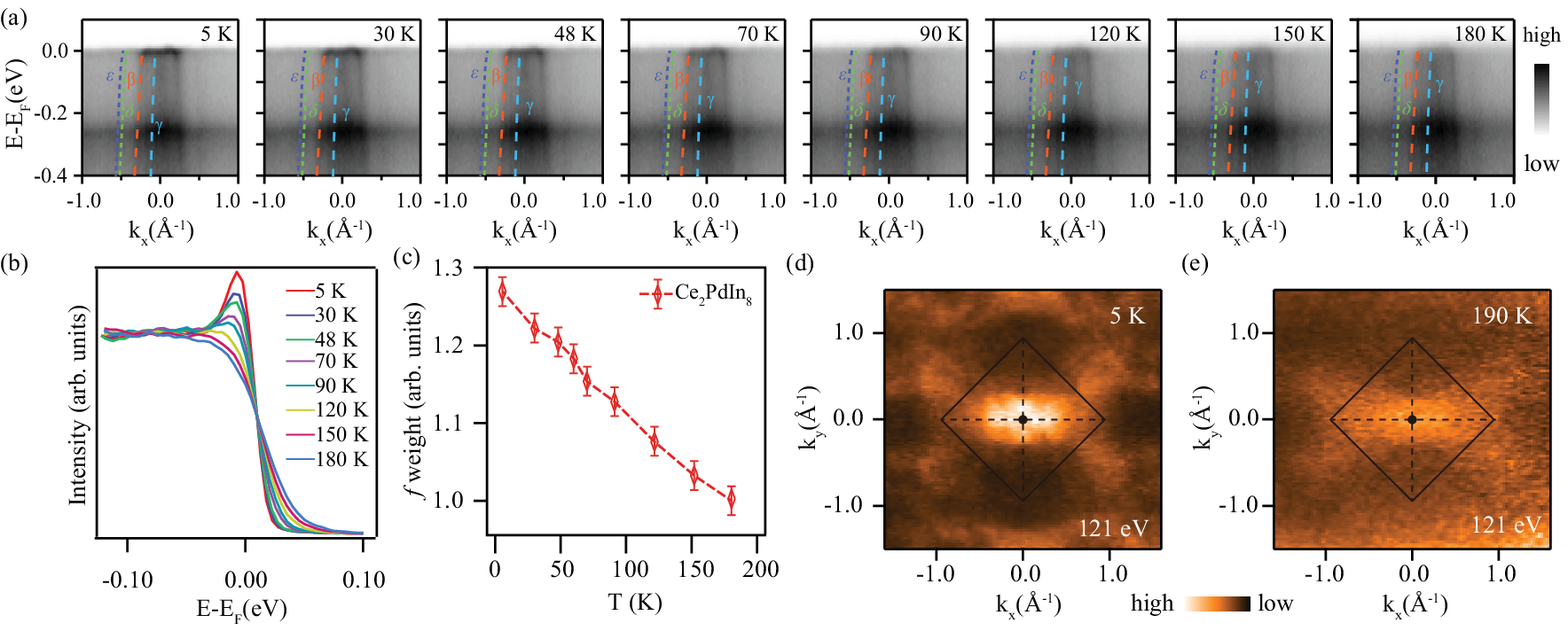}
    \caption{Temperature evolution of the electronic structure of Ce$_2$PdIn$_8$. (a) Resonant (121eV) photoemission intensities along the $M$-$\Gamma$-$M$ direction at different temperatures as labeled, in which the dispersions of conduction bands were extracted by MDCs analysis, see the supplementary material at~\cite{SM}. (b) Temperature dependence of EDCs around ${\Gamma}$. (c) Temperature-dependent quasiparticle spectral weight in the vicinity of $\Gamma$ near $E_F$ for Ce$_2$PdIn$_8$. (d)-(e) Photoemission maps taken at 5~K and 190~K, respectively.}
\label{f electron}
\end{figure*}

Figs.~\ref{bands}(c) and \ref{bands}(d) compare photoemission intensity maps for the $\varGamma$$X$$M$ and $Z$$A$$R$ planes, and their detailed photoemission intensity plots along high-symmetry directions are illustrated in Figs.~\ref{bands}(e-j) using the same colors. The Fermi surfaces in the $\varGamma$$X$$M$ plane consist of a square Fermi pocket around $\varGamma$ and a racetrack pocket centered at $X$ (both formed by the holelike band $\gamma$), and four pockets around $M$ resulting from the Fermi crossings of the electronlike bands $\alpha$, nearly-degenerate $\varepsilon$/$\delta$, and $\beta$. In the $Z$$A$$R$ plane, both the shape of most Fermi pockets and the corresponding band dispersions exhibit only minor variations, confirming the quasi-two-dimensional character of these bands. Still, differences can be resolved between the Fermi surface topologies of these two $k_z$ planes. Changes in both the shape and size of the $\gamma$ pocket in the $Z$$A$$R$ plane imply a three-dimensional character of this band. The main bands in the vicinity of the Fermi level ($E_F$) have similar orbital character to those in the $n$=1 compound CeCoIn$_5$, while band splitting due to interlayer coupling makes the band structure of Ce$_2$PdIn$_8$ more complex. Overall, our experimental data show a qualitative agreement with first-principles calculations, as displayed in Figs.~\ref{bands}(k) and \ref{bands}(l).

In order to highlight the $f$-electron behavior, we chose 121~eV photons (the energy of the Ce 4$d$-4$f$ transition) to realize resonant enhancement of the Ce 4$f$ photoionization cross section.
Figs.~\ref{resonance}(b) and \ref{resonance}(c) show the off-resonance and on-resonance photoemission intensity plots along $\varGamma$-$M$, taken with 114~eV and 121~eV photons, respectively, using $s$-polarized light. The corresponding on-resonance photoemission intensity plot taken with $p$-polarized light is displayed in Fig.~\ref{resonance}(d). Here, $p$- and $s$-polarized light is defined to be from electric fields in and out of the emission plane, which is determined by the analyzer slit and the sample surface normal, as illustrated in Fig.~\ref{resonance}(a).

As shown in Fig.~\ref{resonance}(b), Pd 4$d$ and Ce 5$d$ states dominate the off-resonance spectra, consistent with the data shown in Fig.~\ref{bands}(e). The Ce 4$f$ character is strongly enhanced in the on-resonance cases, as demonstrated by the flat bands in the photoemission intensity plots [Figs.~\ref{resonance}(c) and \ref{resonance}(d)] and the apparent peaks in the corresponding integrated spectra [Fig.~\ref{resonance}(e)]. The three evident features located around $E_F$, -0.25~eV and -2.5~eV below $E_F$ can be well understood by the single impurity Anderson model (SIAM) and were thus assigned as 4$f^1_{5/2}$, 4$f^1_{7/2}$ and 4$f^0$, respectively~\cite{f_electron, CeCoIn5_qiuyun}.
Moreover, we note that the 4$f$ state near $E_F$ shows a sharp peak under $s$-polarized light but a broad hump in $p$-polarized light, and the overall $f$ electron intensity observed with $p$-polarized light is weaker than with $s$-polarized light. Therefore, further investigations of the Ce 4$f$ states in this compound were performed with $s$-polarized light.


Next we performed temperature-dependent ARPES measurements on this compound to investigate the evolution of the itinerant/localized nature of the $f$ electrons with temperature. At 5~K, a flat $f$-electron feature near $E_F$ can be clearly observed in the intensity plot, as demonstrated in Fig.~\ref{f electron}(a). At this temperature, large $f$ spectral weight can be clearly observed near $E_F$,
 suggesting a rather itinerant nature of the $f$ electrons. Upon increasing the temperature, the spectral weight of the $f$ states gradually diminishes while that of the $\beta$ band is still preserved even to rather high temperature (180~K). This temperature dependence is also clearly visible when directly comparing the EDCs at $\Gamma$ taken at different temperatures [Fig.~\ref{f electron}(b)]. We observe a suppression of the $f$ spectral weight upon increasing temperature, and the sharp $f$-electron quasi-particle peak near $E_F$ is no longer visible at 150~K. Fig.~\ref{f electron}(c) presents the temperature dependence of $f$-electron spectral weight in the vicinity of $\Gamma$ in Ce$_2$PdIn$_8$, which is obtained by integrating the EDCs over [$E_F$ - 100 meV, $E_F$ + 10 meV]. Here, the spectral weight has been normalized at 200K and is found to keep arising with the decreasing temperature, showing the same trend as Fig.~\ref{f electron}(b).

For comparison, Fig.~\ref{f electron}(d) and Fig.~\ref{f electron}(e) show the normalized photoemission intensity maps taken at 5~K and 190~K, respectively. In the 5~K Fermi surface map, the normalized intensity around the Brillouin zone center is strongly enhanced, originating from the itinerant $f$ states. In contrast, the $f$ electron spectral weight around the same momentum is greatly suppressed for the map contour taken at 190~K. Note that we have presented both maps using the same color scale, so the much higher spectral weight around the zone center at 5~K should not be artificial. Moreover, as shown in the supplementary material\cite{SM}, direct comparison of MDCs taken at 5~K and 190~K can further confirm the spreading out of the $f$-band spectral weight from the zone center to somewhere with the increasing temperature. All the above data explicitly show the characteristic localized-to-itinerant crossover of $f$ electrons, which has never been reported in the prototypical Ce 218 heavy fermion systems before. Remarkably, the occurrence of hybridization starts from about 120~K in Ce$_2$PdIn$_8$, which is much higher than $T_{coh}\sim 30$~K~\cite{Ce218Kondo1, Ce218Kondo2, Ce218Kondo3, CePdIn_growth}. 

\begin{figure}
    \includegraphics[width=87mm]{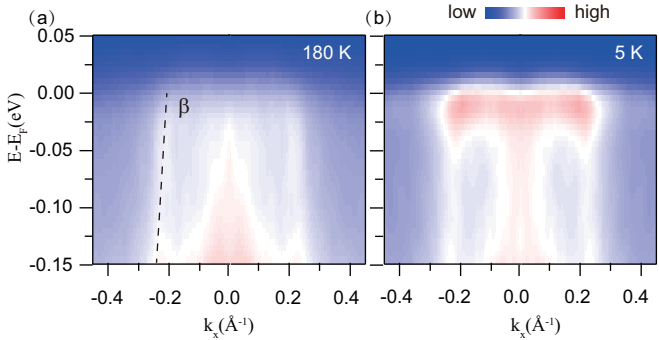}
    \caption{Comparison of the photoemission intensity plot between high and low temperature in Ce$_2$PdIn$_8$. (a) Zoomed-in ARPES data along the $\varGamma$-$M$ direction taken at 180 K. (b) Same as (a), but taken at 5 K.}
\label{comparison}
\end{figure}

In Fig.~\ref{comparison} we present the zoomed-in photoemission intensity plot near $E_F$ to concentrate on the $f$-electron character and quasiparticle bands. Compared with spectra taken at high temperatures, when the $\beta$ band shows linear dispersion, evident bending of this band around $E_F$ can be observed at the low temperature [Fig.~\ref{comparison}(a)]. This is a direct evidence of the hybridization between the $f$ and conduction electrons. According to the standard view on the heavy fermion ground state based on periodic Anderson Model (PAM), the hybridization between the $f$ and conduction electrons would create two separate bands with band bending near $E_F$. Meanwhile, the $f$ spectral weight should be redistributed, with significant enhancement to the ``inside" of the hole-like bands and the ``outside" of the electron-like bands. In Fig. 4(b), we indeed observe the enhancement of the $f$ spectral weight ``inside" of the hole-like $\beta$ band, which is one more evidence for the hybridization between the $f$ and conduction electrons in Ce$_2$PdIn$_8$. Such behavior is similar to that observed in the heavy-fermion compound CeIrIn$_5$ \cite{CeIrIn5_qiuyun}.

\begin{figure}
    \includegraphics[width=\columnwidth]{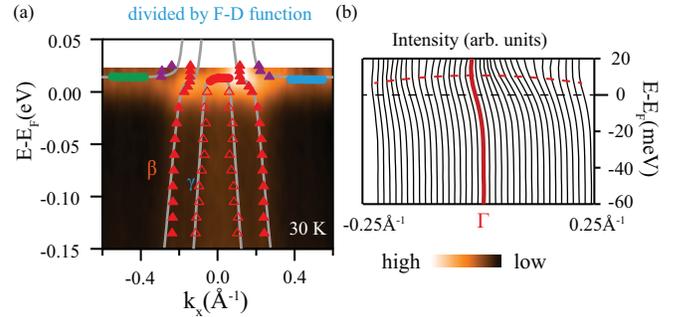}
    \caption{Development of the heavy quasiparticle band in Ce$_2$PdIn$_8$. (a) Photoemission intensity plot along $\varGamma$-$M$ taken at 30 K, divided by the resolution-convoluted Fermi-Dirac distribution. The grey solid lines illustrate how the conduction $\beta$ and $\gamma$ bands hybridize with the $f$ bands. The triangles and solid circles in colors are obtained by fitting individual MDCs and EDCs (see the Supplementary Materials at~\cite{SM}). (b) EDCs corresponding to the spectrum in panel (a) in the momentum window [$-$0.25\,\AA$^{-1}$, 0.25\,\AA$^{-1}$], showing a weakly dispersive feature.}
\label{heavy quasiparticle}
\end{figure}

To further quantitatively investigate the hybridization of conduction bands and $f$ electrons, we zoom in the spectral  feature around $\varGamma$ in the vicinity of $E_F$ and then performed the PAM fitting, which has been proved an effective way to study the $f$ electron behaviors~\cite{CeCoIn5_qiuyun,CeRhIn5_qiuyun}. Details of the band dispersion extraction and the PAM fitting can be found in the supplementary material~\cite{SM}. Here, the resolution-convoluted Fermi-Dirac distribution has been divided in order to probe the band structure slightly above $E_F$, as shown in Fig.~\ref{heavy quasiparticle}(a).

 Quantitatively, a fit to PAM picture gives a renormalized hybridization $V_{k}$=15$\pm$5~meV for both conduction bands, which implies the substantial hybridization strength between $f$ electrons and conduction $d$ bands. Moreover, as shown in Fig.~\ref{heavy quasiparticle}(b), a heavy quasiparticle band is induced with an energy dispersion of around 6 meV, which is larger than that in CeIn$_3$ \cite{zhangyun_CeIn3, Fujimori}, indicating a stronger hybridization in CeIn$_3$.


To summarize, we have characterized the electronic structure and $f$ electron behavior of Ce$_2$PdIn$_8$ using high-resolution resonant and non-resonant ARPES. Besides presenting the low-lying electronic structure of this material, we have shown how localized $f$ electrons in Ce$_2$PdIn$_8$ become partially itinerant and evolve into the heavy-fermion state from a much higher temperature than the coherence temperature $T_{coh}$. 
These findings provide a comprehensive experimental picture of the character of $f$ electrons in Ce$_2$PdIn$_8$, complementing our understanding of $f$ electrons' itinerant-to-localized evolution in the Ce$_n$M$_m$In$_{3n+2m}$ heavy fermion family.

We thank Diamond Light Source for the beam time at beamline I05 (Proposal No. SI16345) that contributed to the results presented here. We grateful acknowledge helpful discussions with Dr. D. C. Peets. The work was supported by the National Science Foundation of China (Grants No.\ 11874330, 11274332, 11574337, 11704073, 11504342, and U1332209), the National Key R\&D Program of the MOST of China (Grants No.\ 2016YFA0300200, 2017YFA0303104). Also, this work is supported by the National Science Centre (Poland) under research grant No. 2015/19/B/ST3/03158 and U1332209), and the Science Challenge Project (No.\ TZ2016004). P.S.'s work at Ames Laboratory (electronic structure calculations) was supported by the U.S. Department of Energy, Office of Basic Energy Sciences, Division of Materials Science and Engineering. Ames Laboratory is operated for the US Department of Energy by Iowa State University under Contract No. DE-AC02-07CH11358. D.W.S is also supported by ``Award for Outstanding Member in Youth Innovation Promotion Association CAS".


\bibliographystyle{apsrev4-1}
%

\end{document}